\documentclass[fp,twocolumn]{jpsj3} %
\usepackage{txfonts}
\usepackage{color}

\usepackage{amsmath}
\usepackage[dvipdfmx]{graphicx}
\usepackage{dcolumn}
\usepackage{bm} 

\title{\boldmath Design of opposed-anvil-type high-pressure cell for precision magnetometry and its application to quantum magnetism}

\author{
{Naoka~Hiraoka}$^{1}$,
{Kelton~Whiteaker}$^{1,2}$,
{Marian~Blankenhorn}$^{3}$,
{Yoshiyuki~Hayashi}$^{1}$,
{Ryosuke~Oka}$^{1}$,
{Hidenori~Takagi}$^{1,3,4}$, and
{Kentaro~Kitagawa}$^{1}$
}

\inst{$^1$Department of Physics, Graduate School of Science, University of Tokyo, Tokyo 113-0033, Japan\\
$^2$Department of Physics and Astronomy, University of British Columbia, Vancouver BC V6T 1Z2, Canada\\
$^3$Institute for Functional Matter and Quantum Technologies, University of Stuttgart, 70569 Stuttgart, Germany\\
$^4$Max Planck Institute for Solid State Research, Heisenbergstra\ss e 1, 70569 Stuttgart, Germany}

\abst{
We have developed a highly sensitive technique to conduct magnetometry under high pressures up to 6.3~GPa using an opposed-anvil type cell, which can detect a weak volume susceptibility as small as $\sim 10^{-4}$.
The high-pressure cell made of non-magnetic binderless tungsten carbide ceramics and CuBe alloy has an optimized geometry to yield a reduced background in the magnetic response, one order of magnitude smaller than those for previously reported high-pressure cells in a commercial SQUID magnetometer.
To further increase the sample-signal-to-background ratio, a conical shaped gasket and cupped anvils are introduced to ensure almost ten times better space efficiency.
The estimate of background contributions is achieved by taking deformation of the cell parts into account.
The magnetization is extracted from the scanned SQUID data using a truncated singular value decomposition (tSVD) linear algebra.
tSVD is shown to give more reliable estimate of magnetization than conventional non-linear least-squares (NLLS) analysis.
The successful application of the new techniques to the measurements of paramagnetic susceptibilities of spin orbit entangled moment under pressures evidences the drastic improvement of their performance.
}

\kword{high pressure, magnetometry, experimental technique, K$_2$RuCl$_6$} 
\begin{document}
\maketitle
\section{Introduction}

Tuning electronic parameters of materials by applying high pressure is useful to explore novel quantum phases and critical phase competitions,
while magnetization is one of fundamental physical quantities to characterize the electronic states.
Bulk magnetometry under high pressure has been extensively utilized to detect the emergence of a superconducting (SC) state or a ferromagnetic (FM) state,
since they are accompanied with an intense negative/positive magnetic signal\cite{FloresLivas2020,Takeda2002}.
In contrast, there have been only a limited number of studies to explore electronic phases with only a weak magnetization, for example, quantum spin liquid.
Only weakly magnetic phases of  interest induced by pressure can be explored in principle, but 
available techniques under a pressure so far are microscopic probes, such as nuclear magnetic resonance (NMR), $\mu$SR, M\"ossbauer spectroscopy, and neutron and X-ray scattering\cite{Biesner2020}.
These probes measure the local internal magnetic fields and provide detailed information on the microscopic state of electrons.
However, they have many constraints.
Observable species of elements for NMR and M\"ossbauer are limited.
$\mu$SR, neutron and X-ray scattering require beam lines and are not suitable for quick screening of materials.
Highly sensitive bulk magnetometry technique that can resolve a weak magnetization is desired for the exploration of unseen electronic states.

High-pressure clamp cell devices that fit to commercial superconducting quantum interference device (SQUID) magnetometers have been widely
 developed in the past two decades\cite{Wang2014,Mito2001,Alireza2009,Kobayashi2007,Tateiwa2011,Giriat2010}, which is quite promising technique to conduct sensitive magnetometry under pressure.
They could achieve a high sensitivity of magnetization as low as $\sim 10^{-6}$~emu, if background signals from a high-pressure cell were negligible.
A high pressure $P$ up to 2 GPa is reachable using piston-cylinder-type cells\cite{Wang2014}.
The maximum pressure of 2 GPa is, however, not enough to induce change in many inorganic materials.
An opposed-anvil-type cell is the only choice to reach a high pressure $P$ beyond 2~GPa inside the very limited space of the SQUID magnetometers.
The available sample space in opposed-anvil-type cell, however, is even smaller compared with those in piston-cylinder-type cell, which reduces the signal-to-background (S/B) ratio substantially.
Recent studies employed opposed-anvil-type cell and succeeded in observing a relatively small---compared with SC/FM cases---
magnetism: an antiferromagnetism of sizably large moments\cite{Giriat2010,Tateiwa2011,Kozlenko2012,Alireza2009},
 and a weak ferromagnetism produced by a non-collinearity\cite{Majumder2018}.
Even with the resolution of magnetization in these studies, it is hard to detect weak signal of $10^{-3}$-$10^{-4}$ emu/mol for quantum magnet with $S=1/2$ moment due to the small volume and hence the small magnitude of the signals.
Further improvement of opposed-anvil-type cell in its S/B ratio is required to conduct the high-pressure magnetomery of weak magnets under high pressures up to several GPa.

In this \textcolor{black}{paper}, we present the refinements of the design of opposed-anvil-type high-pressure cell and of the process of estimating the magnetization, that enable us to study a weak magnetism of quantum magnets.
In section 2, we describe the novel design of high-pressure cell.
By employing a geometrical cancellation approach, we successfully minimized the background magnetization from the cell over the whole temperature and magnetic field ranges.
We then replaced zirconia-based composite ceramics for anvils and piston assemblies with binderless tungsten carbide ceramics with a reduced magnetization.
A conical shaped gasket and cupped anvils are introduced to increase the space efficiency.
In section 3, we present an improved method to extract the sample magnetization from the measured SQUID response curves.
First, we consider the displacement of cell parts upon the application of the pressure explicitly and introduce a new method of background estimation to eliminate the effect of displacement.
Second, we introduce a truncated singular value decomposition (tSVD) linear algebra to extract local magnetizations inside/near the sample room. 
The applications of the developed high-pressure cell and the analysis
 to spin orbit entangled van-Vleck magnet K$_2$RuCl$_6$ are shown.

\section{Improvement of S/B Ratio}
\subsection{Reduction of Background Signals}
\subsubsection{Optimization of Geometrical Cancellation}

\begin{figure}[htbp]
    \centering
    \includegraphics[width=1.0\linewidth]{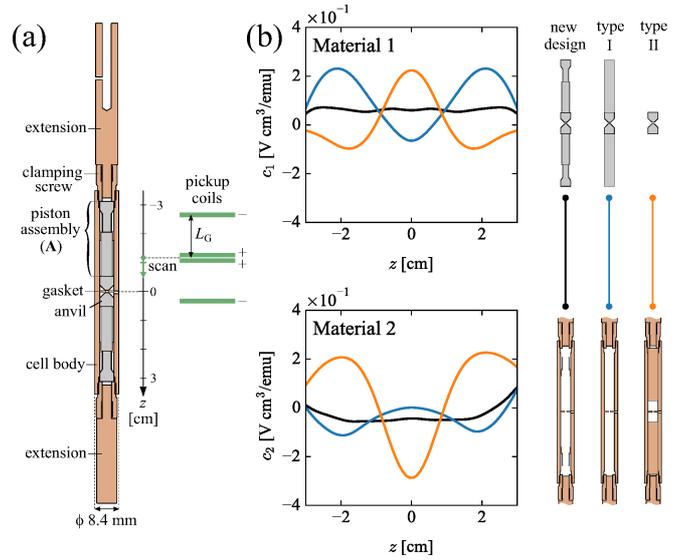}
    \caption{\label{fig:design}{(Color online) Design of high-pressure cell with reduced background signals in a commercial SQUID magnetometer (MPMS XL, Quantum Design).}
    (a) Cross-sectional drawings of the cell assembly. Grey and brown colors indicate the two species of materials, Material 1 and Material 2, respectively. The assembly is placed inside the second-derivative pick-up coils indicated by the green lines during the measurements.
    (b) Simulated partial SQUID scan responses normalized by the volume magnetic susceptibility of the material. The upper (lower) panel is the response $c_1$ and $c_2$ from the parts made of Material 1 or Material 2. The black lines represent responses of the cell with new design. The light blue and the orange lines show responses from the set up with conventional design (type I and type II), where the piston assemblies (part "A" in panel (a)) are replaced by simple columnar pistons made of Material 1 or Material 2, respectively. Cross-sectional views of the cell with the new, the type I and the type II designs are shown next to the graph.
    }
\end{figure}

Our design of high-pressure cell is shown in Fig.~\ref{fig:design}(a).
It follows well-established mechanisms of the opposed-anvil-type cell, consisting of a clamp cell body with locking screws and anvils.
A piston assembly between lock screws and anvils\cite{Kobayashi2007}, and extension\cite{Tateiwa2011} are additionally
 inserted to minimize the SQUID response from the cell ("background signals").
This reduction of response is based on a geometrical cancellation of induction voltages, as we address later.

The array of pickup coils in the commercial SQUID magnetometer (MPMS, Quantum Design), widely used for magnetization measurements, consists of two sets of differential coils with opposite signs (second derivative configuration), which cancels the induction voltage from a far object.
A SQUID response voltage from a point object above and below the pick up coils is proportional to the magnetization of the object\textcolor{black}{. The dependency on $z$, a relative position from the coil center to the object,} is described as\cite{MPMSAN},
\begin{align}\label{eq:response}
    g(z) =& \frac{2}{\left(R^2+z^2\right)^{3/2}}\nonumber\\
    & - \frac{1}{\left(R^2+(z + L_G)^2\right)^{3/2}} - \frac{1}{\left(R^2+(z - L_G)^2\right)^{3/2}},
\end{align}
where $R$ corresponds to the radius of the coils,  $L_G$ is the spacing between the coils.
In the case of Quantum Design MPMS-XL SQUID magnetometer, $R = 0.97$~cm and $L_G=1.519$~cm\cite{MPMSAN}.
Note that a very long homogeneous rod does not contribute to the signal, as it dose not cause any change in magnetic flux in the coils.
This is why the existing high-pressure SQUID cells are quite long along the scanning direction.
Alternatively, one can expect a cancellation of signals from different part of the cell.
$g(z)$ has a positive maximum at $z=0$ and becomes negative when $|z|\gtrsim0.97$~cm.
Careful choice of cell geometry can lead to the cancellation by the contributions from different part of the cell.

The SQUID scan voltage $V(z)$ from a finite size object can be approximated by a convolution of Eq.~\eqref{eq:response} and $z$-slices of the magnetizations $m(z)$:
\begin{align}\label{eq:voltage}
    V(z) = A \int_{-\infty}^{\infty} g(z' \textcolor{black}{-} z) m(z') dz',
\end{align}
where $A$ is a device constant\textcolor{black}{, and the origin of $z$ is at the center of the scan as shown in Fig.~\ref{fig:design}(a)}. 
In the latter part of this paper, we adopt $A$=1~V~cm$^{3/2}$~emu$^{-1}$ so that $V(z)$ is equivalent to the "scaled voltage"  which appears in data files of QD MPMS.
$m(z)$ can be expanded with respect to component materials as
\begin{align}\label{eq:M}
m(z) = \sum_k S_k(z) \chi_k(T, H) H,
\end{align}
where $k$ is an index for material species, $S_k(z)$ is a cross-section of $k$-th material, 
and $\chi_k(T,H)$ is the volume susceptibility of $k$-th material at a temperature $T$ under a magnetic field of $H$.
Then, perfect cancellation is achieved when
\begin{align}\label{eq:intS}
c_k(z) \equiv A \int_{-\infty}^{\infty} g(z' \textcolor{black}{-} z) S_k(z') dz' = 0, 
  \quad\text{for each $k$},
\end{align}
where $c_k(z)$ represents partial SQUID response from objects made of $k$-th material normalized by the volume magnetization.
This gives rise to $V(z) = 0$ at any temperature and field as long as the material is highly homogeneous.
Note that a constant offset during the scan, $V(z) = $const., can be electronically canceled, and does not influence the detection of signal from the setup.
The constraint for zero background response can be relaxed to
\begin{align}\label{eq:intS2}c_k(z) = \text{const.},
    \quad\text{for each $k$}.
\end{align}

\begin{figure}[htbp]
\centering
\includegraphics[width=1.0\linewidth]{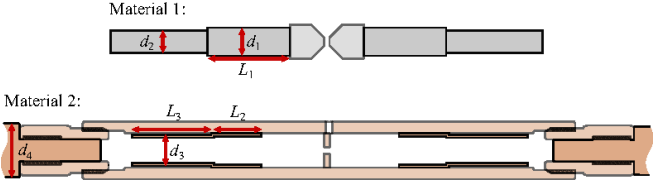}
\caption{\label{fig:optimization}{(Color online) Parameters tuned to optimize the design of the cell.
(upper) Parameters for parts made of Material 1. $d_1(d_2)$: diameters of the inner (outer) rods, $L_1$: lengths of the inner rods. (bottom) Parameters for parts made of Material 2. $d_3$: outer diameters of the outer collars, $d_4$: diameters of the extensions $L_2(L_3)$:lengths of the inner (outer) collars. Discrete values of $L_1$,$L_2$ and $L_3$ with 1 mm steps were tested while $d_1$,$d_2$,$d_3$ and $d_4$ were changed continuously during iterative least-square computations.
The contributions from the other components (anvils, gasket, cell body and screws, shown in light colors) are fixed during the iterations, and was accurately calculated based on the three-dimensional drawings\cite{note1}.}
}
\end{figure}

We design the cell comprising of cylindrical parts as shown in Fig.~\ref{fig:optimization}, modifying the design given by Ref.~\citen{Tateiwa2013}, and conducted a tuning of geometry using eq.~\ref{eq:intS2} as a guide.
The cross-sectional drawing of the cell parts is shown for two materials sectors: Material 1 and Material 2.
Material 1, shown in gray color in Fig.~\ref{fig:optimization}, represents a material with high-compression-strength and high hardness, which is suitable for fabrication of anvils.
Material 2, shown in brown color in Fig.~\ref{fig:optimization} stands for a tough and high-tensile-strength alloy, from which the cell clamp body and screws should be made.
To increase the flexibility of design, we adopted a piston assembly (``A" in Fig.~\ref{fig:design}(a)) composed of inner and outer rods made of Material 1 and their collars made of Material 2, instead of a simple columnar piston.
The choice of materials for pistons was demonstrated to affect the signals from the cell substantially by Tateiwa {\it et al.} \cite{Tateiwa2013}, which motivated us to do further optimization with higher flexibility.
To minimize the SQUID response from the cell in the real set-up, we need to minimize a integral $\int_{-l/2}^{+l/2} |c_k(z)-\langle c_k(z)\rangle |^2 dz$, where $l$=6~cm is a typical scan length and $\langle c_k(z)\rangle $ is an average of $c_k(z)$ in the range of the scan.
The least-square analysis was conducted by optimizing the adjustable parameters of the cell:
the lengths and the diameters of inner and outer rods and their collars and the diameters of extensions.
The shape of the anvils, and the cell body and the screws were fixed, which are shown in light colors in Fig.~\ref{fig:optimization}.
We optimized the cell part made of Material 1 by tuning the three parameters $L_1$, $d_1$ and $d_2$ and the sector made of Material 2 by tuning the four parameters $L_2$, $L_3$, $d_3$ and $d_4$.

Figure~\ref{fig:design}(b) show partial SQUID scan responses of the two materials sectors simulated based on eqs.~\ref{eq:response}-~\ref{eq:intS} with the optimized design in black lines.
For comparison, those with conventional designs only with one-piece cylindrical pistons made of Material 1 (type I) or Material 2 (type II) are also shown in light blue and orange lines, respectively. 
The replacement of the simple pistons with the optimized piston assembly (``A" in Fig.~\ref{fig:design}(a)) suppresses the responses both from Material 1 and Material 2.

\begin{figure}[htbp]
\centering
\includegraphics[width=1.0\linewidth]{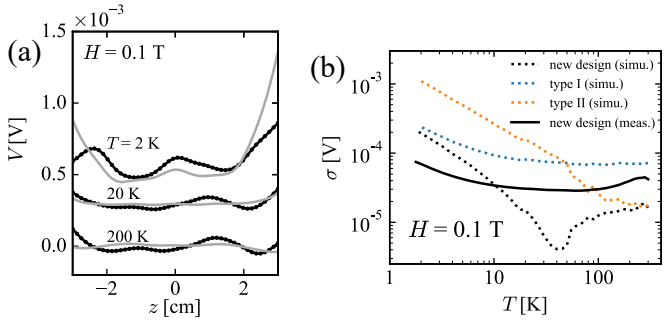}
\caption{\label{fig:Vbg}{(Color online)  Background signals from the cell.
(a) SQUID scan responses of newly designed cell at 2, 20 and 200~K under 1000~Oe obtained by simulations (gray) and by evaluation measurements (black). FCY20A zirconia-based composite ceramic and H17200 Cu-Be alloy were employed as Material 1 and Material 2, respectively (See Fig.~\ref{fig:design} and text). Offsets were added for clarity.
(b) Temperature dependence of the standard deviation in the SQUID scan response of $l$=6~cm, $\sigma$ (solid black line). Simulations for the new and the conventional designs with simple columnar pistons (type I and type II in Fig.~\ref{fig:design}) are also shown for comparison in dashed lines. 
}
}
\end{figure}

\begin{figure}[htbp]
\centering
\includegraphics[width=0.6\linewidth]{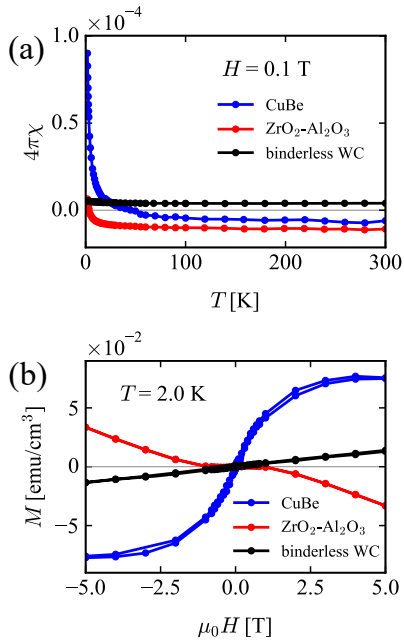}
\caption{\label{fig:materials}{(Color online) The temperature dependence of volume magnetic susceptibilities and magnetization curves of non-magnetic materials: C17200HT Cu-Be alloy, zirconia-based composite ceramic (FUJI DIE Co., Ltd., FCY20A), and binderless WC (NJS Co., Ltd., M78).
}
}
\end{figure}

The whole cell response is accordingly suppressed in comparison with those of conventional design.
This can be verified as shown in Fig.~\ref{fig:Vbg}.
In Fig.~\ref{fig:Vbg}(a), we show simulated and experimental scan voltages of newly designed high-pressure cell at $T$= 2, 20, and 200 K under 0.1 T.
FCY20A zirconia-based composite ceramic of FUJI DIE Co., Ltd. and C17200HT Cu-Be alloy, which has been used in previous high-pressure cells \cite{Kobayashi2007,Tateiwa2011,Tateiwa2013},
were employed as Material 1 and Material 2 to fabricate the cell, respectively, to evaluate our numerical optimization at this stage.
For the simulation, we used experimentally measured $\chi(T)$ of the zirconia-based composite ceramics and Cu-Be alloy shown in Fig.~\ref{fig:materials}(a).
All the Cu-Be alloy parts were silver and/or gold plated after age hardening, to prevent surface oxidation \cite{note2}.
To evaluate the amplitude of the background signal, we \textcolor{black}{employ a standard deviation of $V(z)$ during the entire scan, $\sigma=\sqrt{\frac{1}{n}\sum_{i=1}^{n}(V(z_{i})-\langle V(z_{i}) \rangle )^{2}}$ ($-3 < z_{i} < 3$ cm).} 
Figure~\ref{fig:Vbg}(b) shows temperature dependence of $\sigma$ simulated for the three designs in Fig.~\ref{fig:design}(b) in dashed lines.
The magnitude of background signal from the cell with new design is reduced roughly by a factor of 10 as compared to conventional designs over the whole temperature range,
and thus the concept of optimization using geometrical cancellation is established.
The magnitude of  $\sigma$ and the temperature dependence, which takes a minimum value around several tens K, reproduce the simulated results well, proving the concept actually works.
The tiny deviation of experimental data form the simulation can be ascribed to the heterogeneity of materials.

\subsubsection{Use of New Non-magnetic Material for the cell}
The Material 1 and Material 2 for the cell should have a magnetic susceptibility as small as possible and therefore should be nonmagnetic.
As mentioned in the previous section, Cu-Be alloy and zirconia-based composite ceramics (here after ZrO$_2$-Al$_2$O$_3$) were used in recent studies \cite{Kobayashi2007,Tateiwa2011,Tateiwa2013}.
After the search for materials, we found that binderless tungsten carbide ceramics (here after binderless WC, NJS Co., Ltd., M78) can replace ZrO$_2$-Al$_2$O$_3$ in terms of mechanical properties but with much smaller susceptibility.
Figure 4 shows the comparison of magnetic susceptibility, $\chi$, for nonmagnetic materials, Cu-Be alloy for Material 2 and ZrO$_2$-Al$_2$O$_3$ and binderless WC for Material 1. 
Due to the presence of minor magnetic elements additives, such as Co in Cu-Be alloy or oxygen defects in ZrO$_2$-Al$_2$O$_3$, a profound Curie-like tail at low temperatures is observed for these two materials (see Fig.~\ref{fig:materials}(a))
Binderless WC is definitely different from so-called non-magnetic tungsten carbide ceramics with Ni binder, and include no magnetic additives.
As shown in Fig. 4, the magnetic susceptibility (per unit volume) of this material is much less in magnitude and temperature dependency as compared to that of ZrO$_2$-Al$_2$O$_3$.
Combined with Eqs.~\eqref{eq:voltage} and \eqref{eq:M}, the magnetization data indicates that use of binderless WC instead of ZrO$_2$-Al$_2$O$_3$ as Material 1 reduces the background signals of the cell.
We note that the parts made of binderless WC could be repeatedly used without breaking during actual pressure application measurements, which will be described in detail below.

\subsection{Enhancement of Space Efficiency}
\label{sec:spaceeff}
\begin{figure}[htbp]
    \centering
    \includegraphics[width=1.0\linewidth]{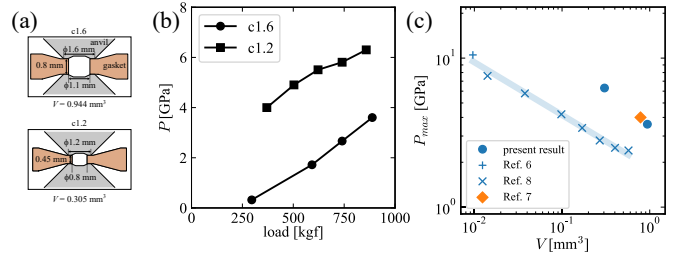}
    \caption{\label{fig:pressure}{(Color online)}
    Cupped binderless WC anvils and conical shaped Cu-Be alloy gasket.
    (a) The cross sections of the gasket and anvils for two culet sizes: "c1.6" and "c1.2".
	(b) Load-pressure curves obtained for the two setups.
	(c) Maximum pressure versus initial sample-space volume.
	 Results of an opposed-anvil-type cell in Ref.~\citen{Alireza2009} and Ref.~\citen{Tateiwa2011} and an indenter cell in Ref.~\citen{Kobayashi2007} are also plotted for comparisons. Cu-Be alloy gaskets are used in Ref.~\citen{Alireza2009}, Ref.~\citen{Tateiwa2011}, while Ni-Cr-Al alloy gasket is used in Ref.~\citen{Kobayashi2007}. The trend for the previous Cu-Be alloy gaskets is indicated in light blue line.
    }
\end{figure}

In addition to the reduction of the background response, we have successfully increased the space efficiency of pressure cell by modifying the design of the gasket and anvils.
We added slanted supports to the gasket, which has been shown to improve the gasket endurance in a high-pressure NMR study\cite{Kitagawa2010}.
The sample space was extended by adding a tapered cavity to the tip of the anvil, and by increasing the relative diameter of the hole of the gasket to that of the tip of the anvil (cullet size).
Cross-sectional views of the newly designed gasket for two types of anvils with a culet size of 1.6 and 1.2~mm are shown in Fig.~\ref{fig:pressure}(a).
Note that conventional techniques utilize a planar gasket and flat tipped anvils.

We tested the pressure efficiency of the newly designed gaskets made of Cu-Be alloy and anvils made of binderless WC using Daphne7373 oil (Idemitsu Kosan Co.,Ltd.) as pressure transmitting medium and a fragment of Pb as a manometer. 
The applied pressure at a low temperature was calibrated by using the pressure dependence of superconducting transition temperature of Pb\cite{Eiling1981}.
The load pressure curves obtained for the two types of anvils are shown in Fig.~\ref{fig:pressure}(b).
The volume of sample space for the culet size of 1.2(1.6)~mm is 0.305(0.944)~mm$^3$ where 0.040(0.092)$\times$2~mm$^3$ comes from cavities of anvils and 0.226(0.760)~mm$^3$ from the hole of the gasket.
The achieved maximum pressure was 6.3(3.6)$\pm$0.3~GPa at 858(888) kgf.

Figure~\ref{fig:pressure}(c) shows maximum generated pressures plotted against the initial volumes of sample spaces for the present setups together with those of existing cells with various culet size of anvils.
For the same culet size of 1.2~mm and the material (Cu-Be alloy) of the gasket, the previous study showed a maximum pressure of 3.4$\pm$0.3~GPa, for a sample space volume of 0.169~mm$^3$. \cite{Tateiwa2011}
Compared with this, the maximum generated pressure and the sample volume space of the present setup are increased by +80\% and +85\%, respectively,
indicating the successful increase of the sample space without sacrificing the maximum pressure.
When our results are compared with the trend line for conventional Cu-Be alloy gaskets  (light blue line in Fig.~\ref{fig:pressure}(c)) at the same maximum pressure, the initial sample space is ten times larger.
It has been reported previously that Ni-Cr-Al gasket can provide a higher space efficiency compared with Cu-Be alloy gasket but suffered from a large background signal and have a small S/B ratio due to its large magnetization\cite{Tateiwa2011}.
The space efficiency and the maximum pressure for the present setup is comparable even to those with Ni-Cr-Al alloy gasket\cite{Kobayashi2007} as shown in Fig.~\ref{fig:pressure}(c).

\section{Improved Method of Analysis}
\subsection{Background under High pressure}
Obtained SQUID scan voltage $V_{\rm total}(z)$ includes contributions from both the sample and the high-pressure cell.
The response of cell, $V_{\rm cell}(z)$, can be measured beforehand and can be subtracted from $V_{\rm total}(z)$ as a background to extract the response from the sample $V_{\rm sample}(z)$.
The application of pressure yields displacement and deformation of the cell parts and $V_{\rm cell}(z)$ is pressure dependent.
The pressure dependence of  $V_{\rm cell}(z)$ gives rise to a substantial error in the estimation of $V_{\rm sample}(z)$ and hence magnetization.
It is, however, not practical to measure the SQUID response from high-pressure cell without sample at each pressure.

\begin{figure}[htbp]
    \centering
    \includegraphics[width=0.8\linewidth]{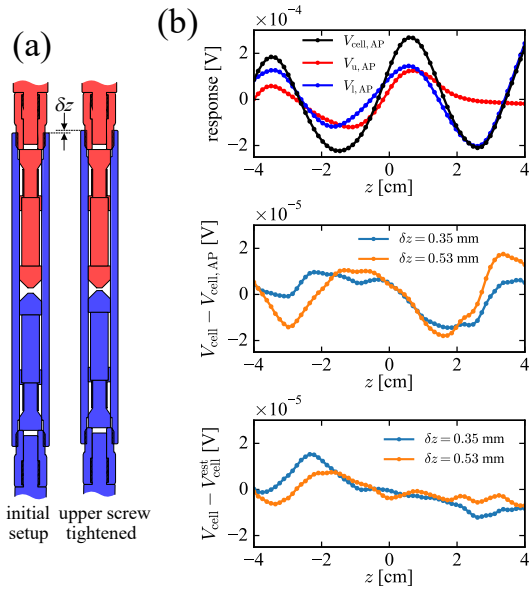}
    \caption{\label{fig:bgsubtraction}{(Color online)
    Experimental SQUID scan voltages of the cell upon screw tightening (clamping).}
    (a) Schematic images of displacements of the cell parts.
    (b) Responses of the whole assembly, upper parts and the other section (upper panel), and the change of the response from whole the cell upon screw tightening (middle panel). The changes in the responses can be compensated by considering the shifts of the lower parts of the cell (bottom panel).  The data shown were acquired at room temperature under 1000~Oe. Anvils made of binderless WC and simple columnar pistons made of Cu-Be alloy was used.  Folded teflon tape was inserted between the anvils instead of a gasket.
    }
\end{figure}

To eliminate the error from the pressure dependence of  $V_{\rm cell}(z)$, we employed a simple method to estimate the effect of deformation and displacement of the cell at each pressure.
We assume that the deformation under pressure takes place only to the gasket inside the cell, when the load is applied to the cell and the upper screw is tightened.
We ignore the deformation in the anvils, pistons, and the others because the stress concentrates on the tips of the anvils and the gasket.
The gasket usually deforms plastically by 50\% or more in height at maximum pressure, while the others are designed to be used well below the elastic limits, $\ll 1$\%.
Then, we can divide $V_{\rm cell}(z)$ into the two contributions from the upper (the red region in Fig.~\ref{fig:bgsubtraction}(a)) and the lower (blue region) parts, and approximate the displacements as a uniform shift of the lower section to the top by $\delta z$.
The upper part in red consists of MPMS rod, the upper extension, the upper screw, the upper piston assembly, and the upper anvil. The lower part in blue consists of the cell body, the gasket, the lower anvil, the lower piston assembly, the lower screw and the lower extension.

\begin{figure}[htbp]
    \centering
    \includegraphics[width=0.5\linewidth]{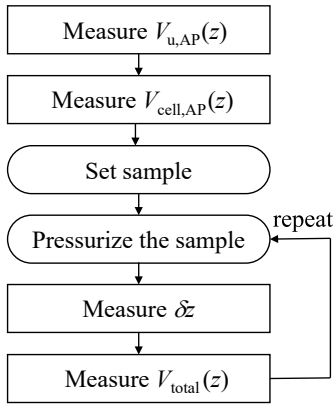}
    \caption{\label{fig:bgsbtscheme}{Measurement scheme to obtain approximated background under high-pressure, $V^{\rm est}_{\rm cell}(z,p)$.}
    }
\end{figure}

We measure \textcolor{black}{two sets of SQUID responses} to estimate the background signals under a pressure with the above approximation: the response from the whole high-pressure cell, \textcolor{black}{$V_{\rm cell,AP}(z)$}; at ambient pressure and that of the upper part, \textcolor{black}{$V_{\rm u,AP}(z)$}\textcolor{black}{, under the same temperature and magnetic field conditions as in the measurement with a sample}.
Then, the SQUID scan voltage of the lower part of the cell, \textcolor{black}{$V_{\rm l,AP}(z)$}, can be calculated as, 
\begin{align}\label{eq:Vlower}
    V_{\rm l,AP}(z) = V_{\rm cell,AP}(z)-V_{\rm u,AP}(z).
\end{align}
Using $V_{\rm u,AP}(z)$ and $V_{\rm l,AP}(z)$, the SQUID scan voltages of the high-pressure cell after pressure application, \textcolor{black}{$V_{\rm cell}(z,p)$}, can be approximated as,
\begin{align}\label{eq:Vallhp}
    V_{\rm cell}(z,p) \simeq V^{\rm est}_{\rm cell}(z,p) &\equiv V_{\rm u,AP}(z)+V_{\rm l,AP}(z+\delta z) \nonumber \\
    & = V_{\rm u,AP}(z)+(V_{\rm cell,AP}(z+\delta z)-V_{\rm u,AP}(z+\delta z)),
\end{align}
where $p$ indicates the applied pressure.
The sequence of measurement is now summarized in Fig.~\ref{fig:bgsbtscheme}.
Figure~\ref{fig:bgsubtraction}(b) show the measured cell responses using a test setup, which indicates that our simple approximation captures the essence of deformation effect under pressure.
We used binderless WC anvils and simple columnar pistons made of Cu-Be alloy, and a folded teflon tape instead of the Cu-Be gasket in the test setup.
The upper parts were fixed to each other using epoxy adhesive, and supported by a plastic straw.
The response from the upper part is shown in the red line in the upper panel of Fig.~\ref{fig:bgsubtraction}(b).
After the measurement, the assembly of the upper parts was inserted into the assembly of lower parts.
The SQUID response from the whole cell with the initial setup is displayed in a black line in the upper panel of Fig.~\ref{fig:bgsubtraction}(b),
which gives an estimate of the contribution from the lower part (blue).
The measured cell response $V_{\rm cell}(z,p)$ and differential responses $V_{\rm cell}(z,p)-V_{\rm cell,AP}(z)$  from the whole cell obtained after the two tightenings, corresponding to the application of pressure, are shown as the light blue and the orange lines in the upper and middle panel of Fig.~\ref{fig:bgsubtraction}(b), respectively.
The distance between the two ends of piston assemblies was measured using a micrometer.
$\delta z$, measured as the reduction of the distance from the initial value, is indicated as the label of each data.
We see that the effect of displacement $\delta z$ of $\sim$0.5 mm, which we observed under a pressure of 3 GPa using gasket for culet size of 1.6~mm (see top panel of Fig.~\ref{fig:pressure}(a)), is as large as $\sim 5\%$ of the signal from the cell.
By compensating the effect of displacement $\delta z$ in the SQUID response from the lower parts (blue line in the upper panel of Fig.~\ref{fig:bgsubtraction}(b)), the background response under a high pressure $V^{\rm est}_{\rm cell}(z,p)$ is estimated:
the deviation $V_{\rm cell}(z,p)-V^{\rm est}_{\rm cell}(z,p)$ is substantially reduced as compared with $V_{\rm cell}(z,p)-V_{\rm cell,AP}(z)$ as shown in the bottom panel of Fig.~\ref{fig:bgsubtraction}(b),
implying that our estimates of $V^{\rm est}_{\rm cell}(z,p)$ can capture well the effect of deformation.
The voltage response $V_{\rm sample}(z)$ from the sample should thus be better tracked by taking $V_{\rm total}(z) -  V^{\rm est}_{\rm cell}(z,p)$ in actual measurements.

\begin{figure}[htbp]
\centering
\includegraphics[width=1.0\linewidth]{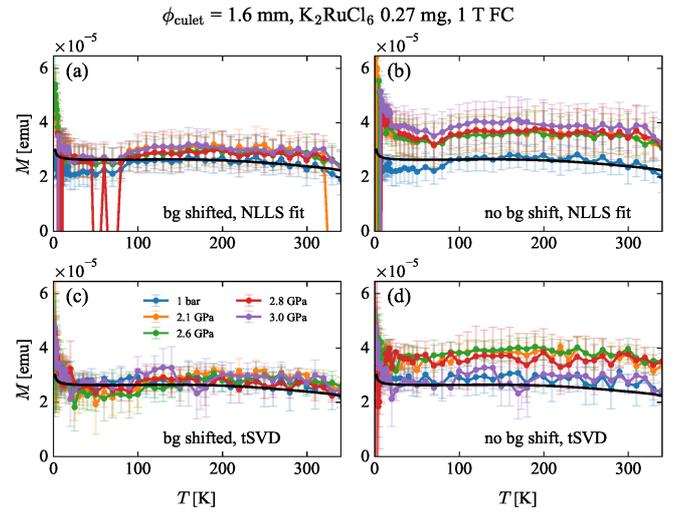}
\caption{\label{fig:hp_mag}(Color online)
$J_{\rm eff}$ van-Vleck magnetism in K$_2$RuCl$_6$ under high pressures, measured with 1.6~mm culet anvils in Fig.~\ref{fig:pressure}
and gasket \cite{note3}.
Magnetization $M$ estimated by a conventional non-linear-least-square fit (a) with and (b) without corrections for the piston displacement,
and $M$ ($\tilde M/\alpha$ in text) processed by truncated SVD method (c) with and (d) without the corrections.
The black lines are ambient-pressure $\chi$ measured without use of a high-pressure cell, shown for comparisons.
$0.27$~mg of sample was used and the low-temperature saturated value of $M=2.5\times 10^{-5}$~emu at 1~T corresponds to
a susceptibility $\chi = 3.5 \times 10^{-3}$~emu/mol, or a volume susceptibility $4\pi\chi = 3.3\times 10^{-4}$.
}
\end{figure}
\begin{figure}[htbp]
\centering
\includegraphics[width=1.0\linewidth]{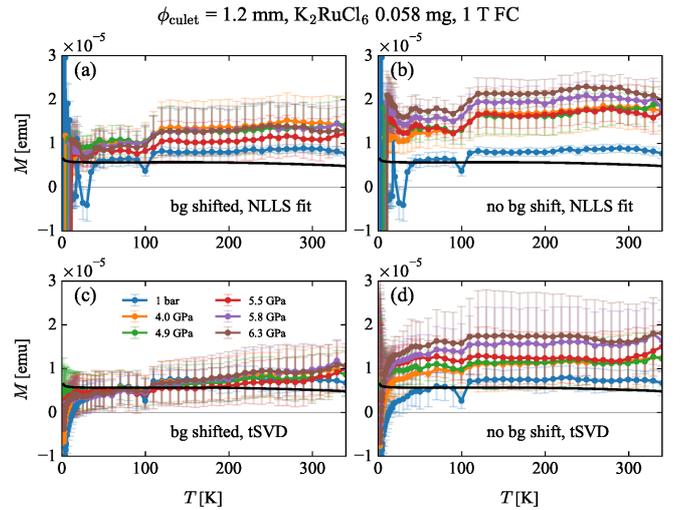}
\caption{\label{fig:hp_mag_c1_2}(Color online)
Measurement for K$_2$RuCl$_6$ with the smaller 1.2~mm culet anvils
and the gasket in Fig.~\ref{fig:pressure}.
$0.058$~mg of sample was used for this case. See text and Fig.~\ref{fig:hp_mag} for the corrections and the black lines.
}
\end{figure}



To demonstrate the effect of background subtraction method in actual magnetization measurements, we show the result of magnetization measurements under various pressures for van-Vleck magnetism of Ru$^{4+}$ pseudospin $J_{\rm eff}=0$ state for K$_2$RuCl$_6$ in Figs.~\ref{fig:hp_mag} and ~\ref{fig:hp_mag_c1_2}.
A conventional approach to extract a magnetic moment of a given object is a non-linear least-squares (NLLS) fitting to the scanned SQUID voltages.
The NLLS fitting introduces four parameters\cite{MPMSAN}: sample magnetization $M_{\rm sample}$,  the displacement of sample from the origin $z_0$, and the two parameters $a_0$ and $a_1$ representing the $z$-dependent instrumental voltage offsets, $a_1 z + a_0$.
To eliminate the background contributions, as we stated above, $V^{\rm est}_{\rm cell}(z,p)$  must be subtracted from $V_{\rm total}(z)$ prior to fitting.
The magnetization data in the panels (a) and (b) are obtained by the NLLS method after the background subtraction
with (a) and without (b) the correction for the cell deformation according to Eq.~\eqref{eq:Vallhp}.
The measurements in Fig.~\ref{fig:hp_mag} was conducted with a larger sample mass of 0.27~mg for $\phi$1.6 mm culet and those in Fig.~\ref{fig:hp_mag_c1_2} were measured for 0.058~mg sample for $\phi$1.2 mm culet. 
Figure~\ref{fig:hp_mag} shows better statistics apparently because of the large sample mass and hence signal than those in Fig.~\ref{fig:hp_mag_c1_2}.
The ambient-pressure susceptibility measured without cell is shown by the black line.
The low-temperature saturated value is $\chi = 3.5 \times 10^{-3}$~emu/mol or a volume susceptibility $4\pi\chi = 3.3\times 10^{-4}$ at 1 T.
Since the van-Vleck magnetism of ionic Ru$^{4+}$ in perfect cubic crystal field is determined by the spin-orbit coupling $\lambda$, $\chi$ is expected to be almost independent of $P$. 
In the data in Figs.~\ref{fig:hp_mag}(b) and \ref{fig:hp_mag_c1_2}(b) without the correction of deformation while the ambient pressure data agree reasonably with the data taken without the cell (black solid line), the data under pressures show significant deviation from those without the cell.
The deviation very likely represent the change of the background due to the deformation induced by the pressure.
Once the correction of the deformation is introduced in Figs.~\ref{fig:hp_mag}(a) and \ref{fig:hp_mag_c1_2}(a), however, the deviation from the black solid line under pressures is substantially reduced.
This evidences our approach to compensate the effect of deformation is working well.

\subsection{Estimate of Magnetization using Truncated Singular Value Decomposition}
When signals from the sample are not large enough as compared with the background, the NLLS procedure fitting does not give a good estimate of magnetization of sample $M_{\rm sample}$.
It is clear from the comparison \textcolor{black}{between} Fig.~\ref{fig:hp_mag}(a) with a large amount of sample and Fig.~\ref{fig:hp_mag_c1_2}(a) with a small amount of sample.
We cannot eliminate the errors completely in the estimate of $V^{\rm est}_{\rm cell}(z,p)$ based on Eq.~\eqref{eq:Vallhp}. 
This comes mainly from deformations far from the center, errors in the length of belt-drive actuator, or variation in the initial positions of the cell for different setups.
The estimation of magnetization should be further improved if we can eliminate these effects far away from center reasonably.
There is another issue in the conventional NLLS process.
The parameter $z_0$, inevitable mechanical offset from the center, makes the fitting non-linear, requiring numerical iterations to solve it.
$|z_0|$ is typically not more than a few millimeters, while a fitted value of $M_{\rm sample}$ is very sensitive to $z_0$.
When the preprocessed voltage $V(=V_{\rm total}(z) - V^{\rm est}_{\rm cell}(z,p) - a_1 z \textcolor{black}{-} a_0)$ does not have similarity to $g(z)$ around the sample position, the NLLS procedure fails in finding an appropriate $z_0$ for $A  M_{\rm sample} g(z-z_0)$ in $V$.
The resultant $z_0$ and hence $M_{\rm sample}$ are sometimes unreliable, as seen by frequent data jumps over error bars in Figs.~\ref{fig:hp_mag}(a) and \ref{fig:hp_mag_c1_2}(a).
It may then worth introducing an alternative scheme, being more robust against the inevitable errors in $V^{\rm est}_{\rm cell}(z,p)$.

\textcolor{black}{We propose a new method to integrate the local magnetization around the origin, where the sample is located within the integration area, instead of finding $z_0$.
It is accomplished by inferring $m(z)$, the local magnetization from the $z$-slices.}
$m(z)$ needs to be deconvoluted somehow from $V$, but another advantage of doing so is that errors of background signals outside the integration area can be effectively eliminated which merits in avoiding the errors in $V^{\rm est}_{\rm cell}(z,p)$ far away from the center.
By substituting the integral in Eq.~\eqref{eq:voltage} by a summation over the data range, typically $-3$ to $3$~cm, and assuming $m(z)$ beyond the range to be zero,
$m(z)$ might be directly solved as a vector $m$;
\begin{align}\label{eq:illposed}
    m = G^{-1} V,
\end{align}
where a vector $V$ is a set of preprocessed voltages after a linear least-square fit to $a_1 z + a_0$, and $G$ is a matrix form of $g(z'-z)$\textcolor{black}{: $G_{ij} = Ag(z'_i - z_j)$}.
This kind of inverse transform is generally known as an ill-posed problem.
Eq.~\eqref{eq:illposed} is unsolvable because the determinant of $G$ is zero.

To give a good approximation for the vector $m$, however, we employed a regularization by truncated singular value decomposition (tSVD)\cite{ILLPOSED,Hansen1994}, 
which is the simplest method among known regularization techniques.
Since the subtracted data often contain sizable errors of background estimation, the instrumental white noise is not the main source of voltage deviations.
Other regularization techniques like Tikhonov regularization is not a good candidate here because an unphysical extra parameter is required.

\begin{figure}[htbp]
\centering
\includegraphics[width=1.0\linewidth]{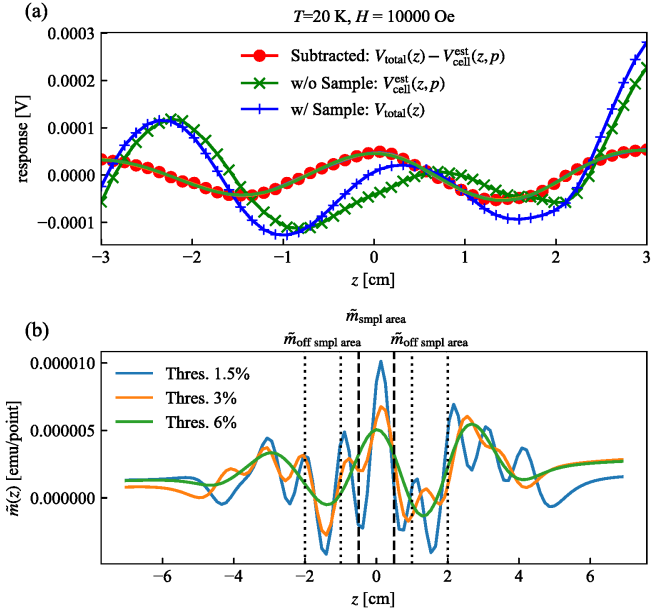}
\caption{\label{fig:tsvd}(Color online)
\textcolor{black}{Typical processed result by the tSVD analysis.  The data set at 2.1~GPa and 20~K in Fig.~\ref{fig:hp_mag}(c) is used. (a) Voltage signals, before tSVD, $V_{\rm total}(z) - V^{\rm est}_{\rm cell}(z,p)$, $V^{\rm est}_{\rm cell}(z,p)$, $V_{\rm total}(z)$ are shown. The traces in front of red points are inversely converted $V$, $=\sum_j G_{ij} \tilde m_j$, from the result of tSVD. The difference between red points and the traces represents a side effect of the truncation, but is negligibly small even for larger threashold value for the truncation.
 (b)  $z$-sliced magnetization $\tilde m_j$ obtained for the several choices of the truncation threashold.
}
}
\end{figure}

Assuming the white noises are small enough (compared with the errors from the background estimate), an experimental observation puts constrains during the approximation as,
\begin{align}\label{eq:constraint}
    V_i =\sum_j G_{ij} m_j.
\end{align}
Here $i$ is taken within the acquisition area $-3{\rm ~cm} < z_i < +3{\rm ~cm}$, but $j \,(> i)$ runs beyond that and we took \textcolor{black}{$-7{\rm ~cm} < z_j < +7{\rm ~cm}$}.
$m$ beyond the acquisition region is required because the cell height plus significant region in the function $g$ is much larger than 6~cm.
By the use of SVD, a least-square result is obtained so that $\sum_j m_j^2$ is minimized subject to the constraints in Eq.~\eqref{eq:constraint}.
First, SVD decomposes the non-square matrix $G$, as $G = u \Sigma v^{\rm T}$,
 where $u$ and $v$ are unitary matrices and $\Sigma$ is a diagonal matrix of singular values $\lambda_k$.
 A pseudo inverse of $G$ is then approximated to
\begin{align}
\tilde G^{\#} = v \tilde \Sigma^{\#} u^{\rm T}.
\end{align}
$\tilde \Sigma^{\#}$ is a diagonal matrix of inverse singular values  $\lambda_k^{-1}$, after performing an appropriate cutoff for a small $\lambda$ 
and reducing a rank of $\tilde \Sigma^{\#}$ accordingly.
\textcolor{black}{Namely, $\lambda_k^{-1}$ $(k > k_{\rm max})$ in $\tilde \Sigma^{\#}$ are replaced by zeros.}
Thus, a least-square approximation for the vector $m$ becomes
\begin{align}\label{eq:tildem}
\tilde m = v \tilde \Sigma^{\#} u^{\rm T} V.
\end{align}
Finally, the magnetic moment at the sample position is estimated as,
\begin{align}\label{eq:msmpl}
\tilde M =\sum^{\rm sample~area} \left\{ \tilde m  - \langle \tilde m \rangle_{\rm off~sample~area}\right\}.
\end{align}
$ \langle \tilde m \rangle_{\rm off~sample~area}$ is an average of $\tilde m$ outside the sample area to subtract a constant offset in $\tilde m$.
This subtraction is necessary because of huge numerical errors on a constant due to a low-ranked fashion of $G$, 
where $\int^{\infty}_{-\infty} g(z) = 0$ for the second-order coil arrangement. 
We use $|z_i| < 0.5$~cm for sample area and $1 < |z_i| < 2$~cm for $\langle \tilde m \rangle_{\rm off~sample~area}$.
\textcolor{black}{We choose 3\% of the largest signular value $\lambda_1$  as a threshold to construct $\tilde \Sigma^{\#}$, and its rank is reduced to 10 ($=k_{\rm max}$).
Selection of a smaller threshold value means that $\tilde m$ can reproduce $V$ more precisely.
However, experimental errors in $V$ or insufficient data length for $V$ tend to produce spiky glitches and/or huge vibrations in $\tilde m$.
A larger threshold value in turn spoils $z$-resolution of $\tilde m$, and thus, the moderate truncation is prefered.
Example plots of the deconvoluted $\tilde m$, for the measurement of K$_2$RuCl$_6$, are shown in Fig.~\ref{fig:tsvd}.
}

Three advantages of the tSVD approach in comparison to standard NLLS are there:
first, only magnetic moments near the center can be extracted, as we already stated;
second, Eqs.~\eqref{eq:tildem} and \eqref{eq:msmpl} have linear forms so that $\tilde M$ is always expected to be proportional to the actual $M$,
 that is, $\tilde M = \alpha M_{\rm sample} +  M_{\rm unresolved~err~from~bkg}+  M_{\rm noise}$. 
The last two terms are expected to distribute around zero.
Then, we can obtain the true $M_{\rm sample}$ value if we take an average of $\tilde M/\alpha$ by many sets of measurement.
A constant $\alpha$ may deviate from unity because of the cutoffs in $\tilde \Sigma^{\#}$,
 and needs to be calibrated by test measurement or numerical analysis, $\alpha = 1.13$ for the parameters used above;
third, since $\tilde M$ is a summation near the center, there is no need to fit a possible $T$-dependent center position change, as long as it is within the sample area.
From these reasons, we conclude that this technique is more reliable than NLLS in particular when $M_{\rm sample}$ is small compared with the background.

Figures~\ref{fig:hp_mag}(c,d) and Figures~\ref{fig:hp_mag_c1_2}(c,d) show the results solved by the tSVD scheme applied to the same SQUID data used for the NLLS fits in the panels (a,b).
The error bars represent standard deviations in $\tilde m_{\rm off~sample~area}$.
The results with the correction of background shown in Fig.~\ref{fig:hp_mag}(c) and Fig.~\ref{fig:hp_mag_c1_2}(c) demonstrate that the erroneous pressure dependence is appreciably reduced as compared with those from the NLLS analysis, in particular in Fig.~\ref{fig:hp_mag_c1_2}(c) with a small signal from the sample.
This clearly indicates that the tSVD results with these high-pressure setups are more reliable.
The deviations during the test measurements are as small as $5\times 10^{-6}$~emu.
This is good enough to trace quantum magnetism of $S=1/2$ spin system.
Note that the Curie paramagnetsim of $S=1/2$ at 300~K is of the order of $\sim10^{-5}$~emu when measured with 1~$\mu$mol of spins at 1~T.


To summarize, we have developed successfully a precision magnetometry under a high pressure up to 6.3~GPa which demonstrated to detect weak volume susceptibilities of $3.3\times 10^{-4}$  and resolve magnetism of paramagnet.
The key ingredients are the successful improvement of design of the high-pressure cell, which gives an order of magnitude smaller background signal from the pressure cell as well as higher space efficiency than the cells with conventional design, and the introduction of a more reliable estimate of background signal from the cell and an analysis method.
In this study, the device was installed in a commercial SQUID magnetometer which
allows us rapid evaluation of magnetization under a pressure between 1.8 and 400~K.
The lowest temperature of 1.8~K, which can be easily generated by liquid $^4$He, might be low enough to study quantum magnetism of many strongly correlated $d$-electron systems.
To investigate $f$-electron materials with much lower energy scale of magnetic interactions, however, magnetization measurement down to a much lower temperature than 1.8~K is required.
A high-pressure SQUID magnetometry in a $^3$He refrigerator \cite{iQuantum,Sato2013} or a $^3$He-$^4$He dilution refrigerator is obviously a promising direction of our technique in the future.

\begin{acknowledgments}
We thank Y.~Uwatoko, N.~Tateiwa, K.~Matsubayashi, H.~Takahashi, and Y.~Tsuyuki for discussion and experimental support.
N.~Tateiwa and K.~Matsubayashi gave us detailed usage hints for high-pressure SQUID measurements.
The binderless WC material was introduced to us by Y.~Uwatoko.
We also particularly thank Kathrin Pflaum and staffs in Max-Planck-Institute machineshop for fabrication of the pressure cell.
This work was supported partly by Japan Society for the Promotion of
Science (JSPS) KAKENHI Nos. 19H01836 and 17K14335, and partly by JSPS Core-to-Core Program (A) Advanced Research Networks.
\end{acknowledgments}
\bibliography{document} 

\begin{thebibliography}{10}
\bibitem{note1}
$S_k(z)$ is automatically computed from an output of 3D CAD. This is done by
  sumations at the respective $z$ slices: $\oint (x, y, 0)/2 \cdot d {\mathbf
  n}$, where $n$ is a normal vector of a mesh triangle given in STEP file
  format.
\bibitem{note2}
The thickness of plating silver or gold is typically micrometers and its effect
  on SQUID response is negligible.
\bibitem{note3}
The gasket used here is a prototype, and the volume of initial sample room was
  0.75 times of that in the final design shown in Fig.5.
\end{thebibliography}

\end{document}